\documentclass[runningheads]{llncs}

\usepackage[utf8]{inputenc}
\usepackage{amssymb}
\usepackage{amsmath}
\usepackage{amsthm}
\usepackage{mathtools}
\usepackage[linesnumbered,ruled,vlined]{algorithm2e}
\usepackage{tikz}
\usepackage{bm}
\newcommand{\inv}{^{-1}}
\usepackage{thmtools}
\usepackage{thm-restate}
\newcommand{\R}{\mathbb{R}}

\newcommand{\polylog}{\textnormal{polylog}}
\newcommand{\poly}{\textnormal{poly}}
\usetikzlibrary{trees}
\usepackage{biblatex}
\title{Quantum-Inspired Classical Algorithm for Slow Feature Analysis}
\author{Daniel Chen \inst{1} \and
Yekun Xu\inst{2} \and
Betis Baheri \inst{3} \and 
Samuel A. Stein \inst{5} \and
Chuan Bi \inst{4} \and 
Ying Mao \inst{5} \and 
Qiang Guan \inst{3} \and 
Shuai Xu \inst{1} } 

\authorrunning{D. Chen et al.}

\institute{Case Western Reserve University, Cleveland OH 44106, USA \\
\email{\{txc461, sxx214\}@case.edu} \and
Florida International University, Maimi FL 33199, USA \\
\email{yxu040@fiu.edu} \and
Kent State University, Kent OH 44240, USA \\
\email{\{bbaheri, qguan\}@kent.edu} \and 
National Institute on Aging, National Institute of Health, Baltimore MD 21224, USA \\
\email{chuan.bi@nih.gov} \and
Fordham University, The Bronx NY 10458 \\
\email{\{sstein17, ymao41\}@fordham.edu}}
\date{}

\begin{document}
\maketitle

\begin{abstract}
    Recently, there has been a surge of interest for quantum computation for its ability to exponentially speed up algorithms, including machine learning algorithms. However, Tang suggested that the exponential speed up can also be done on a classical computer. \cite{tang2019quantum} In this paper, we proposed an algorithm for slow feature analysis, a machine learning algorithm that extracts the slow-varying features, with a run time $O(\polylog(n)\poly(d))$. To achieve this, we assumed necessary preprocessing of the input data as well as the existence of a data structure supporting a particular sampling scheme. The analysis of algorithm borrowed results from matrix perturbation theory, which was crucial for the algorithm's correctness. This work demonstrates the possible application and extent for which quantum-inspired computation can be used.
\end{abstract}

\section{Introduction}
There has been a great uprising of interest in quantum computation. By exploiting superposition and entanglement, two main properties of quantum computation that cannot be reproduced on classical computers, quantum algorithms can accomplish tasks polynomial, or even exponentially faster than its classical counterpart. Recently, Google's demonstrated quantum supremacy; their quantum processor was able to complete tasks that were estimated to take an infeasible amount of time for a classical computer in minutes. \cite{arute2019quantum} This physical realization of the power of quantum computation motivates the possibility of quantum computation becoming the main mode of computation in the future.

Knowing the potential benefits of quantum computation, an immediate problem researchers try to tackle is to improve the run time of machine learning algorithms. Machine learning has been proven to be invaluable in practically every field. However, as the size of the problem increase, some machine learning algorithms become infeasible to implement. Thus, most machine learning algorithms would significantly benefit from the reduced run time quantum computation offers. One of the most influential works in the field of quantum machine learning is the quantum algorithm developed by Harrow, Hassidim, and Lloyd for solving linear system of equations, or sometimes known as the HHL algorithm. \cite{harrow2009quantum} The algorithm provides a method of solving linear systems of equations in poly-logarithmic time with respect to the number of variables and the conditional number (the ratio between the largest and smallest singular value), an exponential speed up to the classical solution. However, the theoretical success shed lights on the yet mature state of quantum computation. Many more problems ought to be addressed prior to implementing the HHL, or more generally, quantum machine learning algorithms. \cite{aaronson2015read, biamonte2017quantum} For example, many quantum algorithms require a quantum-state preparation assumption, which is based on quantum random access memory (QRAM) \cite{giovannetti2008quantum} and/or some further assumptions on the data structure \cite{kerenidis2016quantum}. Thus, the current state of quantum computation, both hardware and software, is not yet ready for implementation of complex machine learning algorithms.

To compensate for the absence of quantum computation, there is a rise in quantum-inspired methods, which is a term vaguely referring an imitation of quantum-like properties on classical computers. One of the breakthroughs of quantum-inspired algorithms was done by Tang: a sublinear-time algorithm for recommendation systems. \cite{tang2019quantum} On the high level, this was accomplished through sampling rows and columns of matrices to approximate linear algebraic operations without explicitly constructing the matrix. This innovation lead to the emergence of many other quantum-inspired machine learning algorithms that are all able to achieve sublinear time complexity. \cite{tang2018quantum, chia2018quantum, gilyen2018quantum, ding2019quantum} This breakthrough suggests that classical computers can offer the speed of quantum algorithms only polynomially slower as oppose to exponential. This raises interesting questions regarding the theoretical properties of machine learning algorithms and quantum computation. Recently, Chepurko et. al has reinterpreted this quantum-inpsired technique as leverage or ridge leverage score sampling and greatly improved the bounds on the run time, making the method even more comparable to that of a quantum computer. \cite{chepurko2020quantum} Also recently, Gilyen, Song, and Tang combined the quantum-inspired method with stochastic gradient descent to solve the problem of linear regression. \cite{gilyen2020improved} This technique also achieved a run time closer to that of the quantum algorithm for regression. These two examples shows that the field of quantum-inspired (machine learning) algorithm holds great potential in replicating quantum run time in classical computers.

In this paper, we focused on the problem of slow feature analysis. As its name suggests, it is an algorithm for extracting the slow-varying features from a fast-varying input signal. The ability to find the time invariant features allow us to identify the characteristics of the environment or about the presence of an entity. One can intuitively think about it as looking at a child jumping up and down. The visual signals we receive are rapidly changing and the image ends up blurry. However, most people would not have trouble recognizing that the blurry object is, in fact, a child. The time-invariant features of the child allows us to identify the object even under sub-optimal circumstances. Slow feature analysis, in essence, attempts to extract those features that changes slowly. 

This article presents the quantum-inspired version of slow feature analysis. Using a data structure assumption analogous to the quantum state preparation, we were able to achieve a run time that is poly-logarithmic to the input size and at most quadratic to the dimension of the data set. In proving the correctness of the algorithm, we borrowed techniques from matrix perturbation theory, which gave us the ability to establish an upperbound for singular vectors of a perturbed matrix. Our algorithm shows the potential for quantum-inspired methods to be applied onto a wide variety of machine learning algorithms.

\subsection{Related Work}

Slow feature analysis was first developed by Wiskott and Sejnowski \cite{wiskott2002slow} in 2002, and since then, the concept has been applied to a variety of problems. Naturally, this algorithm provides significant insight in the field of computer vision. Zhang and Tao was the first to apply slow feature analysis to the problem of human action recognition and found great potential \cite{zhang2012slow}. Recently, the concept of slow feature analysis was applied onto an unsupervised deep learning model to analyze multi-temporal remote sensing images \cite{du2019unsupervised}. 

It also has great implications regarding theoretical neuroscience. The neurons of the primary visual cortex is classified into simple cells and complex cells, and slow feature analysis resembles much of the physiological functions of complex cells \cite{berkes2005slow}. By viewing slow feature analysis in a more probabilistic perspective also helps integrate the algorithm into statistical representations of the visual cortex, which is a popular way of representing neural function \cite{turner2007maximum, fiser2010statistically}. 

A quantum version for slow feature analysis was developed by Kerenidis and Luongo \cite{kerenidis2018quantum}. They assumed a QRAM model for retrieving data as well as data preprocessing, namely nonlinear expansion and differentiation. At the end, they achieved a run time complexity dependent only on the norm of the input, the conditional number of the input, and specified error parameters. Experiments were also done comparing the quality of the quantum algorithm and its classical counterpart. By reinterpretting slow feature analysis into a dimension reduction algorithm, they tested for classification accuracy in the embedded space using MNIST. The results show that the quantum algorithm is comparable to the classical one and can be used as a good tool for multiclass classification problems. 

\subsection{Our Methods and Results}

In this paper, we adopted methods from the breakthrough in quantum-inspired methods first developed by Tang \cite{tang2019quantum} and later by Chia et al. \cite{chia1910sampling} for the problem of slow feature analysis. This method was largely inspired by Frieze, Kannan, and Vempala with their sublinear time algorithm for low rank approximation. \cite{frieze2004fast} The sublinear time was achieved through a norm-based sampling procedure. More concretely, given a vector $x \in \R^n$, and let $x_i$ denote the $i$-th entry and $\|x\|$ denote the $\ell_2$ norm of $x$. We can define a distribution over $x$ where the index $i$ is outputted with probability $|x_i|^2/\|x\|$. By repeatedly sampling from vectors/matrices, we can obtain approximate forms of the vectors and matrices without explicitly performing linear algebraic operations. 

In analyzing the correctness of our main result, we begin by analyzing the run time of the algorithm. Then, we go onto analyzing the error bound. The difficulty of this procedure comes in the process of proving the correctness. Due to the probabilistic nature of the algorithm, each step was built off of the previous error-prone steps. Most procedures can be bounded by exploiting properties of norms. However, the effect of performing singular value decomposition on non-exact matrices requires matrix perturbation theory. Davis-Kahan sin$\theta$ theorem \cite{stewart1990matrix} provides a great tool for doing so. This technique can also potentially provide theoretical guarantees for approximating various other procedures such as data centering. The error analysis is followed by a suggestion for picking the parameters for each step for customization. And finally, we finish by giving an alternative algorithm for querying the desired matrix rather than obtaining samples. 

Through the methods described above, we obtain the following result stated as a theorem.

\begin{restatable}[]{thm}{Main}
\label{eq:MainTheorem}
Given the data matrix $X \in \R^{n \times d}$ and the differentiation of $X$, denoted by $\dot X$, as well as the sample and query access of the transpose of the two matrices given by the data structure in Lemma \ref{eq:dataStructure}. We can perform Slow Feature Analysis to obtain the ability to sample and query from $Y \in \R^{n \times J}$, the projection of the data onto the slowest varying subspace, in time complexity
\begin{align*}
    O(\polylog(n), \poly(d, J, \|X\|_F, \|\dot X\|_F, \frac{1}{\epsilon}, \frac{1}{\theta}, \frac{1}{\gamma}, \frac{\|\dot X\|}{\|X\|}))
\end{align*}
with an additive error $\epsilon$ and a constant success probability.
\end{restatable}

\section{Preliminary}
\subsection{Notation}
For a vector $x \in \R^n$, $x(i)$ denotes the $i$-th entry of $x$. Similarly, for a matrix $A$ in $\R^{m\times n}$, $A(i,j)$ denotes the $(i,j)$-th entry of the matrix, $A(i,\cdot)$ denotes the $i$-th row, and $A(\cdot, j)$ denotes the $j$-th column. In this article, we only consider the $\ell_2$-norm of a vector, which we denote as $\|x\|$. For matrix norms, $\|A\|$ denotes the spectral norm and $\|A\|_F$ denotes the Frobenius-norm. We use $I$ to denote the identity matrix, which has $1$ on its diagonals and $0$ elsewhere.

For any matrix $A \in \R^{m \times n}$, the singular value decomposition (SVD) is defined to be the decomposition into $U\Sigma V^T$, where $U$ and $V$ are unitary matrices, meaning $U^TU = I$ and $V^TV = I$, and $\Sigma$ is a diagonal matrix. For simplicity, we use $ u_i$ and $v_i$ to denote the $i$-th column of $U$ and $V$ respectively, and $\sigma_i$ as the $i$-th entry, and by convention, the diagonal entries are sorted in a non-increasing order. We call $u_i$ the left singular vectors, $v_i$ the right singular vectors, and $\sigma_i$ the singular values.

The notion of \textit{approximate isometry} will be used frequently in the algorithms later. It is defined as follow.
\begin{definition}[Approximate Isometry]
A matrix $U \in \R^{m\times n}$ is said to be an $\alpha$-approximate isometry if $\|U^T U - I \| \leq \alpha$.
\end{definition}
Intuitively speaking, an approximate isometry is a matrix that is close to an unitary matrix. Settling for the approximate matrix enables the faster runtime of our algorithm. 

For a non-zero vector $x \in \mathbb{R}^n$, to sample from $x$ means to draw an index $i \in [n]$ following the distribution 
\begin{align}
    \mathcal{D}_x (i) = \frac{x_i^2}{\|x\|^2}
\end{align}
For a non-zero matrix $A \in \mathbb{R}^{n\times d}$, we define the $n$-dimensional vector $\Tilde{A} = \begin{pmatrix} \|A(1,\cdot)\|^2 & \|A(2, \cdot)\|^2 & \dots & \|A(n,\cdot)\|^2 \end{pmatrix}^T$. Thus, the sampling distribution for a matrix is defined by $\mathcal{D}_{\tilde A}$.

\subsection{Slow Feature Analysis}
Given an $d$-dimension input time-series vector $\bm{x}(t) = \begin{bmatrix} x_1(t) & x_2(t) & \dots & x_d(t) \end{bmatrix}^T$ with $t \in [ t_0, t_1]$. Let $h: \R^d \to \R^m$ be a function that nonlinearly expands a $d$-dimensional signal to a $m$-dimensional one. Let $\bm{z}(t) = {h(\bm x(t))}$. The goal of the algorithm is to find a $J$-dimensional vector $\bm{y}(t)$, where $y_j(t) = \bm{w}_j^T \bm{z}(t)$, such that 
\begin{align}
    \Delta (y_j) = \langle \dot{y_j}^2 \rangle = \bm{w_j}^T \langle \dot{\bm{z}}\dot{\bm{z}}^T \rangle \bm{w_j} 
\end{align}
is minimized. The notation $\dot f$ denotes the derivative with respect to time. The operation $\langle f \rangle$ on some function $f$ is defined as the time average, $ \frac{1}{t_1 - t_0} \int_{t_0}^{t_1} f(t) dt$. There are a few constraints to satisfy: 

\begin{enumerate}
\item $\langle y_j \rangle = 0$
\item $\langle y_j^2 \rangle = 1$
\item $\forall j' < j, ~ \langle y_j y_{j'} \rangle = 0$
\end{enumerate}

Assume that the expanded signal $\bm{z}(t)$ has 0 mean and unit covariance. Then, it can be shown that the problem reduces to an eigenvalue problem and the constraints will be automatically satisfied when the weight vectors are orthonormal. 

The rough diagram of the problem is shown as follows:
\begin{align*}
\bm x(t) = \begin{pmatrix} x_1(t) \\ x_2(t) \\ \vdots \\ x_d(t) \end{pmatrix} \xrightarrow[Transform]{Nonlinear} \bm z(t) = \begin{pmatrix} z_1(t) \\ z_2(t) \\ \vdots \\ z_m(t) \end{pmatrix} \xrightarrow{find} \bm y(t) = \begin{pmatrix} y_1(t) \\ y_2(t) \\ \vdots \\ y_J(t) \end{pmatrix} = \begin{pmatrix} \bm w_1^T \bm z(t) \\ \bm w_2^T \bm z(t) \\ \vdots \\ \bm w_J^T \bm z(t) \end{pmatrix}
\end{align*}

The algorithm given by Wiskott and Sejnowski \cite{wiskott2002slow} is given in Algorithm \ref{alg:SFA}.

\begin{algorithm}\label{alg:SFA}
\caption{Standard Algorithm for Slow Feature Analysis}
\KwIn{Signal $\bm x(t)$}
\KwOut{Signal $\bm y(t)$ where minimal velocity variance}
Normalize the input signal, $\bm{x}(t)$, such that it has $0$ mean and variance $1$. \\
Apply a nonlinear expansion ${h}(\bm{x}) = \bm{z}(t)$. For example, a quadratic expansion will result in $\bm{z}(t) = \begin{bmatrix} x_1(t) & \dots & x_I (t) & x_1(t) x_1(t) &x_1(t)x_2(t) $ \dots $ x_I(t)x_I(t) \end{bmatrix}^T$ \\
Normalize $\bm{z}(t)$ such that it has $0$ mean and unit covariance (this step is called \textit{sphering} or \textit{whitening}). In other words, we want $\langle \bm{z} \rangle = 0$, and $\langle \bm{zz^T} \rangle = \bm{I}$. \\
Apply PCA to the matrix $\langle \bm{\dot{z}\dot{z}^T} \rangle$\\
Out put signal $\bm y(t)$, where $y_n(t) = \bm w_j^T \bm z(t)$, where $\bm w_j$ is the eigenvector corresponding to the $j$-th smallest eigenvalue we got from the previous step. 
\end{algorithm}

Our work was inspired by Kerenidis and Luongo's work on quantum slow feature analysis \cite{kerenidis2018quantum}. They have adopted slow feature analysis to solve classification problems, which was originally proposed by Berkes. \cite{berkes2005slow} The reformulation of the problem is as follow.

Given $n$ data points, $x_i \in \R^d$, $i \in [n]$, each belonging to one class $C_k$, $k \in [K]$. We still want to obtain the weight $w_j$ such that 
\begin{align}
    \Delta(y_j) = \frac{1}{\sum_k{{|C_k|}\choose{2}}} \sum_{k = 1}^K \sum_{s,t \in C_k, s < t} (w_j^T x_s - w_j^T x_t )^2
\end{align}
is minimized, subjecting to the constraint that 
\begin{enumerate}
    \item $\frac{1}{n} \sum_{k=1}^K \sum_{i \in C_k} w_j^T x_i = 0$
    \item $\frac{1}{n} \sum_{k=1}^K \sum_{i \in C_k} (w_j^T x_i)^2 = 1$
    \item $\frac{1}{n} \sum_{k=1}^K \sum_{i \in C_k} (w_j^T x_i)(w_{j'}^T x_i) = 0, ~\forall j \neq j'$.
\end{enumerate}

In this interpretation, the temporal aspect of the problem is modified to the changes within each class of label. Thus, the differentiated matrix, $\dot X \in \R^{\mathcal{C} \times d}$, where $\mathcal{C} = \frac{1}{\sum_k{{|C_k|}\choose{2}}}$, is defined to be the matrix such that each row stores the difference between two distinct data points in the same class.  Similar to the classic slow feature analysis above, the data can be non-linearly expanded if appropriate. An algorithm of slow feature analysis for classification is described in Algorithm \ref{alg:SFAclass}.

\begin{algorithm}\label{alg:SFAclass}
\caption{Algorithm for Slow Feature Analysis for Classification}
\KwIn{The normalized data matrix $X \in \R^{n \times d}$}
\KwOut{Output matrix $Y \in \R^{n \times J}, J = K - 1$}

Find the covariance matrix $B = X^T X$ \\
Differentiate the data matrix $X$ to get $\dot X$ \\
Find the differentiated whitened data matrix $\dot Z = \dot X B^{-1/2}$ \\ 
Find the $J$ slowest eigenvector of $\dot Z$ to get $W$ \\
Return $Y = \dot ZW$, the projection of the data onto $W$.
\end{algorithm}

One downside to this interpretation is that the size of the derivative matrix grows quadratically to the number of inputs, which may be computationally expensive. Thus, a common solution is to sample a random subset for differentiation.\cite{berkes2005slow} However, this interpretation allows for the interchangability of differentiation and whitening of the data, which was already implemented in step 3 of Algorithm \ref{alg:SFAclass}.\cite{kerenidis2018quantum} This avoids having to construct the whitened derivative matrix, which would take at least linear time with respect to the number of data. 

\subsection{Linear Algebra via Sampling}

We assume the existence of a special data structure to support sampling. This data structure assumption has been used in various other places \cite{chia2018quantum, kerenidis2016quantum, tang2019quantum, tang2018quantum}. The properties of the data structure is described as follow: 

\begin{lemma} \label{eq:dataStructure}
There exists a data structure storing a non-zero vector $v \in \mathbb{R}^{n}$ with $O(n \log n)$ space that does:
\begin{enumerate}
    \item $O(\log n)$ time entry-wise query and updating
    \item $O(1)$ time output of the $\ell_2$ norm of $v$
    \item $O(\log n)$ time sampling from $\mathcal{D}_v$
\end{enumerate}
\end{lemma}

\begin{lemma}
Let $A \in \mathbb{R}^{n \times d}$ be a non-zero matrix, $\Tilde{A} \in \mathbb{R}^n$ where the $i$-th entry stores $\|A(i,\cdot)\|^2$. Then, there exists a data structure storing $A$ in $O(nd \log nd)$ space, supporting
\begin{enumerate}
    \item $O(\log nd)$ time entry-wise query and updating
    \item $O(1)$ time output of $\|A\|_F^2$
    \item $O(\log n)$ time output of $\|A(i,\cdot)\|^2$
    \item $O(\log nd)$ time sampling from $\mathcal{D}_{\tilde{A}}$ or $\mathcal{D}_{A(i,\cdot)}$
\end{enumerate}
\end{lemma}

The notion of sample and query access is further generalized to arbitrary vectors and matrices. It turns out that given the ability to sample and query from matrices, we can gain the sample and query access to other matrices without explicitly constructing the matrices resulted from intermediate operations. The generalization was done by Chia et. al \cite{chia1910sampling}.

\begin{definition}
For a vector $v \in \mathbb{R}^n$, we have $Q(v)$, \textit{query access} to $v$ if for all $i \in [n]$, we can obtain $v(i)$. Similarly, we have query access to matrix $A \in \mathbb{R}^{n\times d}$ if for all $(i,j) \in [n]\times [d]$, we can obtain $A(i,j)$.
\end{definition}

\begin{definition}
For a vector $v \in \mathbb{R}^n$, we have $SQ_\nu (v)$, sample and query access to $v$ if we: 
\begin{enumerate}
    \item can perform independent samples from $v$ following the distribution $\mathcal{D}_v$ with expected cost $s(v)$
    \item have query access $Q(v)$ with expected cost $q(v)$
    \item obtain $\|v\|$ to multiplicative error $\nu$ with success probability $9/10$ with expected cost $n_\nu (v)$
\end{enumerate}
\end{definition}

\begin{definition}
For a matrix $A \in \mathbb{R}^{n \times d}$, we have $SQ_{\nu_1}^{\nu_2}$, sampling and query access to to $A$ if we:
\begin{enumerate}
    \item have $SQ_{\nu_1}(A(i,\cdot))$, for all $i \in [m]$, as described in Definition 2.2 with cost $s(A)$, $q(A)$, $n_{\nu_1}(A)$ respectively
    \item can sample from $\Tilde{A}$ following the distribution $\mathcal{D}_{\Tilde{A}}$ with cost $s(A)$
    \item can estimate $\|A\|_F^2$ to multiplicative error $\nu_2$ in cost $n^{\nu_2}(v)$
\end{enumerate}
\end{definition}

So, for example, if $n$-dimensional vector is stored in the data structure described in Lemma 1, $s(v) = q(v) = O(\log n)$. Furthermore, for some vector $v$, we denote $sq_\nu(v) = s(v) + q(v) + n_\nu(v)$. Similarly, we denote $sq_{\nu_1}^{\nu_2}(A) = s(A) + q(A) + n_{\nu_1}(A) + n^{\nu_2}(A) = O(\log nd)$, if stored in the data structure described in Lemma \ref{eq:dataStructure}.

\section{Proof}
This section presents the proof of our main result, which is stated as a theorem as follow. The algorithm for Theorem \ref{eq:MainTheorem} is presented in Algorithm \ref{alg:MainAlgo}.
\Main*

\begin{algorithm}\label{alg:MainAlgo}
\caption{Quantum-Inspired Slow Feature Analysis}
\KwIn{data matrix $X^T$, normalized and polynomially expanded (if needed), ${\dot{X}^T} \in \R^{d \times \mathcal{C}}$, obtained through differentiating $X$}
\KwOut{sample from $Y(j, \cdot)$, for some specified $j$} 
Gain sample and query access to $\hat U^T \in \R^{r \times n}, \hat V^T \in \R^{r \times d}, \hat \Sigma\inv \in \R^{r \times r}$, the approximate SVD of $X$ \\
Gain sample and query access to $\hat Z^T = \hat V \hat U^T$ \\
Gain sample and query access to $\widehat {B^{-1/2}} = \hat V \hat \Sigma\inv \hat V^T$ \\
Gain sample and query access to the product $\hat {\dot Z} = \dot X \widehat{B^{-1/2}}$  \\
Gain sample and query access to right singular vectors of $\hat {\dot Z}$ \\
Store the lowest $J$ right singular vectors of ${\dot Z}$ in matrix $\hat W$ \\
Output a sample from $Y(j,\cdot)$ by computing $Y =  \hat Z(j,\cdot) \hat W$
\end{algorithm}

This section will be divided into four parts. In section 3.1, we will analyze the overall run time of the algorithm. In section 3.2, we will show the upperbound for the error produced throughout. Following that, section 3.3 gives suggestions as to how to pick the parameters to bound the overall error into one the user desire. And finally, in section 3.4, we will introduce a slight modification on Algorithm \ref{alg:MainAlgo} that supports query access.

\subsection{Run time analysis}

The runtime of each step, and hence the algorithm, follows from the lemmas and corollaries from section 3. We denote the runtime of step $i$ as $RT_i$. Furthermore, unless otherwise specified, we will denote the parameters picked for the $i$-th step with subscripts $i$ as well. 

The first step involves estimating the SVD of a matrix, which involves the following lemma.

\begin{lemma}[\cite{chia1910sampling}]\label{alg:SVD}
Let $A \in \R^{m \times n}$. Given $SQ(A)$, $SQ(A^T)$, a singular value threshold $\sigma > 0$, and error parameters $\epsilon, \eta \in (0,1)$, there exists an algorithm which gives $D$ and a succinct description of $\hat{U}$ and $\hat{V}$ with probability 9/10 in time complexity $O(\frac{ \|A\|^{24}}{\epsilon^{12} \sigma^{24} \eta^6}sq(A))$. The matrices $\hat{U}, \hat{V}, D$ satisfies that $\hat{U} \in \mathbb{R}^{m \times r}, \hat{V}^{n \times r}$ are $O(\eta \epsilon^2)$-approximate isometries, $D$ is a diagonal matrix. And $r = O(\frac{\|A\|_F^2}{\sigma^2 (1 -\eta)^2})$. We also obtain $SQ_{\eta\epsilon^2}^{\eta\epsilon^2}(V)$ where $s(\hat{V}) = O(\frac{\|A\|_F^{18}}{\epsilon^8 \sigma^{18}\eta^4})$ and $q(\hat V) = O(\frac{\|A\|_F^8}{\epsilon^4 \sigma^8 \eta^2})$.
\end{lemma}
Thus, the run time of step 1 follows straightforwardly as presented in Lemma \ref{alg:SVD}. 
\begin{align}
    RT_1 = O(\frac{\|X\|_F^{24}}{\epsilon_1^{12}\sigma^{24}\eta_1^6 } \log nd)
\end{align}
and we get that $SQ(\hat U) = SQ(\hat V) = O(\frac{\|X\|_F^{18}}{\epsilon_1^8 \sigma^{18}\eta_1^4})$. Furthermore, we store $\hat \Sigma\inv$ in the data structure described in Lemma \ref{eq:dataStructure}, which takes $O(r\log r)$, where $r = O(\frac{\|X\|_F^2}{\sigma^2 (1-\eta_1)^2})$. This provides $SQ(\hat \Sigma\inv) = O(\log r)$.

The second step involves an approximate matrix multiplication, which is described in the following lemma. 

\begin{lemma}[\cite{chia1910sampling}] \label{MatrixMult}
Let $A \in \mathbb{R}^{n \times d}$, $B \in \mathbb{R}^{d \times p}$, and $\epsilon, \delta \in (0,1)$ be error parameters. We can get a concise description of $U \in \mathbb{R}^{m \times t}, D \in \mathbb{R}^{t \times t}, V \in \mathbb{R}^{p \times t}$ such that with 9/10 probability $\|AB - UDV\|_F \leq \epsilon$, where $t = O(\|A\|_F^2\|B\|_F^2/\epsilon^2)$. Furthermore, we get $sq(U) = O(t^2 sq(A^T)), sq(V) = O(t^2sq(B))$. Thus, we can compute a succinct description of $UDV$ in time $O(t^3 + \frac{t^2}{\delta^2}(sq(A^T) + sq(B)))$, and obtain $SQ(UDV)$ where $sq(UDV) = \Tilde{O}(t^2 \max(sq(U),sq(V))).$
\end{lemma}

Thus, the runtime of step 2 follows from Lemma \ref{MatrixMult} with samples and queries from $\hat U$ and $\hat V$. Let $t = O(\frac{\|\hat U\|_F^2 \|\hat V\|_F^2}{\epsilon_2^2})$, and we get
\begin{align}
    RT_2 &= O(t^3 + \frac{t^2}{\delta_2^2}SQ(\hat U)) \\
    &= O(\frac{r^4\|X\|_F^{18}}{\epsilon_1^8 \epsilon_2^4 \sigma^{18}\eta_1^4 \delta_2^2})
\end{align}
We also get $SQ(\hat Z) = \tilde O (\frac{r^4\|X\|_F^{18}}{\epsilon_1^8 \epsilon_2^4 \sigma^{18}\eta_1^4})$.

For step 3, no additional computation is needed to gain sample and query access to $\widehat{ B^{-1/2}}$. The form $\hat V \hat \Sigma\inv \hat V^T$ supports the sampling procedure well, it can be done through the following lemma.

\begin{lemma}[\cite{tang2018quantum}]\label{alg:MatrixVec}
Given $SQ(V^T), Q(w)$ where $V \in \mathbb{R}^{n\times k}$ and $w \in \mathbb{R}^k$, we can get $SQ_\nu(Vw)$ with (expected) time complexities $q(Vw) = O(kq(V)+q(w))$, $s(Vw) = O(k \mathcal{C}(V,w)(s(V) + kq(V)+ kq(w)))$ where \begin{align}
    \mathcal{C}(V,w) = \frac{\sum_i w(i)^2 \|V(\cdot,i)\|^2}{\|\sum_i w(i)V(\cdot,i)\|^2}
\end{align}
\end{lemma}

Using this lemma, we can obtain queries in $O(rsq(\hat V)) = O(\frac{\|X\|_F^{20}}{\epsilon_1^8 \sigma^{20}\eta_1^4(1-\eta)^2)}$ time and sample access in $O(r^2 sq(\hat V)) = O(\frac{\|X\|_F^{22}}{\epsilon_1^8 \sigma^{22}\eta_1^4(1-\eta)^4})$ time. More details are provided in the proof of theorem 3.1 in \cite{chia1910sampling}.

The runtime of step 4 also follows directly from Lemma \ref{MatrixMult}. Let $k = O(\frac{\|\dot X\|_F^2 \|B^{-1/2}\|_F^2}{\epsilon_4^2}) = O(\frac{\|\dot X\|_F^2}{\epsilon_4^2\theta^2})$. 
\begin{align}
    RT_4 &= O(k^3 + \frac{k^2}{\delta_4^2}SQ(B^{-1/2})) \\
    &= O(\frac{\|\dot X\|_F^2\|X\|_F^{22}}{\epsilon_1^8\sigma^{22}\eta_1^4(1-\eta)^4\epsilon_4^2\theta^2\delta_4^2})
\end{align}
where $\theta$ denotes the smallest singular value of $X$. Furthermore, we get $SQ(\hat {\dot Z}) = O(\frac{\|\dot X\|_F^2\|X\|_F^{22}}{\epsilon_1^8\sigma^{22}\eta_1^4(1-\eta)^4\epsilon_4^2\theta^2})$. 

For step 5, we follow Lemma \ref{alg:SVD}. We choose our singular value threshold as the minimum singular value for ${\dot Z}$, which we will denote by $\gamma$, and get the following run time. 
\begin{align}
    RT_6 &= O(\frac{\|\hat {\dot Z}\|_F^{24}}{\epsilon_5^2 \gamma^{24}\eta_5^6}SQ(\hat {\dot Z})) \\
    &= O(\frac{\|X\|_F^{22}\|\dot X\|_F^{26}}{\epsilon_1^8\sigma^{22}\eta_1^4(1-\eta)^4\epsilon_4^2\epsilon_5^2 \gamma^{24} \theta^{26}\eta_5^6})
\end{align}
And we get the query access to the right singular vectors of $\hat A$, which outputs each entry in $O(\frac{\|\hat {\dot Z}\|_F^8}{\epsilon_4\gamma^8 \eta_5^2}) = O(\frac{\|\dot X\|_F^8}{\epsilon_4\gamma^8 \theta^{8} \eta_5^2})$. 

Step 6 requires no further computation since we gained query access to $\hat W$ already in the previous step. One would store the matrix into the data structure described in Lemma \ref{eq:dataStructure} with $\tilde O(\frac{dJ\|\dot X\|_F^8}{\epsilon_4\gamma^8 \theta^{8} \eta_5^2})$ time for $O(\log dJ)$ sample and query access later. 

To output a sample from the $j$-th row of $\hat Y$, we follow the procedures in Lemma \ref{alg:MatrixVec}, and it takes $O(d^2 \frac{\|\dot X\|_F^8}{\epsilon_4\gamma^8 \theta^{8} \eta_5^2})$ time. Again, $C(Z, W(\cdot, i)) = O(1)$ since $Z$ is approximately orthogonal. 

Thus, we see that the overall algorithm runs in logarithmic dependency on $n$ and polynomial dependency on $d$, as well as a few parameter values, the minimum singular value of $X$, and the norm of $X$ and $\dot X$.

\subsection{Correctness}

For the error analysis, we would go through the error produced in each step by following the notation. For some matrix $A$, we denote the theoretical true value with no alternations on the notation, $A$ suffices. Suppose the previous step for computing $A$ is prone to error, which represents most but not all quantities, we denote $\tilde A$ for the theoretical value obtained assuming there is no error in the current step. And lastly, we use $\hat A$ to denote the actual value obtained from the algorithm. Most of the error bounds follow the argument below:
\begin{align}
    \|A - \hat A\| = \|A - \tilde A + \tilde A - \hat A\| \leq \|A - \tilde A \| + \|\tilde A - \hat A\|
\end{align}
where the expression above works for any norm or operations that satisfies the triangle inequality. Furthermore, all notations from the previous section is adopted for this section, and we denote $E_i$ for the error resulted from step $i$ to simplify the notation. 

Step 1 of the algorithm computes an approximate SVD for the data matrix $X$ with Lemma \ref{alg:SVD}. However, we will reference Tang's work \cite{tang2018quantum}, which slightly modifies upon Lemma \ref{alg:SVD}, for implementation and error bound. The consequence of the modification is as follow.

\begin{corollary}[\cite{tang2018quantum}]\label{alg:PCA}
Given $SQ(A), \sigma, k, \eta$ with the guarantee that for $i \in [k]$, $\sigma_i \geq \sigma$ and $\sigma_i^2 - \sigma_{i+1}^2 \geq \eta\|A\|_F^2$, then, there exists a procedure such that the matrix $\hat V$ the algorithm in Lemma \ref{alg:SVD} outputs satisfies $\| V - \hat V\|_F \leq \sqrt{k}\epsilon$ for some $\epsilon \in (0, 0.01)$. Furthermore, $|\hat \sigma_i - \sigma_i| \leq \frac{\eta}{10} \|A\|_F$.
\end{corollary}

We will discuss the specifics of the procedure in the section \ref{subsec:parameter}. Thus, by Corollary \ref{alg:PCA} we have that $\|U - \hat U \|_F \leq \sqrt{r}\epsilon_1$, so is $\|V - \hat V\|_F$. 

Step 2 of the algorithm requires performing an approximate matrix multiplication, which implements the following lemma \ref{MatrixMult}. Given $\hat U, \hat V$, we want to consider the difference $\|UV^T - \widehat{UV^T} \|_F$ through the following computation. 
\begin{align}
    \|UI_{n,d}V^T - \widehat{UV^T} \|_F &\leq \|UI_{n,d}V^T - \hat U \hat V^T \|_F + \|\hat U \hat V^T - \widehat{UV^T}\|_F \\ 
    &\leq \|UI_{n,d}V^T - \hat U I_{r,d}V^T\|_F + \|\hat U I_{r,d}V^T\hat U \hat V^T \|_F + \epsilon_2 \\ 
    &\leq \|V^T\|\|UI_{n,d} - \hat U I_{r,d})\|_F + \|\hat U\|\|I_{r,d}V - \hat V\|_F + \epsilon_2 \\
    &\leq \sqrt{d}\epsilon_1(1 + \eta_1\epsilon_1^2) + \epsilon_2 \label{eq:E2}
\end{align}

For step 3 of the algorithm, we can query and obtain access to $\widehat{B^{-1/2}}$ with the existing sample and query access in an error free way. So, we want $\|V(\Sigma^T\Sigma)^{-1/2}V^T - \widehat{V(\Sigma^T\Sigma)^{-1/2}V^T}\|_F$.
\begin{align}
    &\|V(\Sigma^T\Sigma)^{-1/2}V^T - \widehat{V(\Sigma^T\Sigma)^{-1/2}V^T}\|_F \\
    &\leq \|V(\Sigma^T\Sigma)^{-1/2}V^T - V(\Sigma^T\Sigma)^{-1/2}\hat V^T\|_F + \|V(\Sigma^T\Sigma)^{-1/2}\hat V^T - \hat V \hat \Sigma^{-1} \hat V^T\|_F + \epsilon_3 \\
    &\leq \|V(\Sigma^T\Sigma)^{-1/2}\|\|V - \hat V\|_F + \|\hat V\|_F (\|V(\Sigma^T\Sigma)^{-1/2}\| + \|\hat V\|\|\hat \Sigma\|) + \epsilon_3 \\
    &\leq \sqrt{r}\epsilon_1/\theta + (1 + \sqrt{r}\epsilon_1)(2 + \sqrt{r}\epsilon_1)/\theta + \epsilon_3
\end{align}

Step 4 of the algorithm is another matrix-multiplication. Again, using Lemma \ref{MatrixMult}, we wish to bound $\|\dot X B^{-1/2} - \widehat{\dot X B^{-1/2}}\|_F$.
\begin{align}
    \|\dot X B^{-1/2} - \widehat{\dot X B^{-1/2}}\|_F &\leq \|\dot X B^{-1/2} - \dot X \hat B^{-1/2} \|_F + \epsilon_4 \\
    &\leq \|\dot X\|E_3 + \epsilon_4
\end{align}

Step 5 of the algorithm requires slightly more nuance. The difficulty comes with justification for performing SVD on a matrix with error. We wish to bound the difference between the right singular vectors of the theoretical matrix and the perturbed matrix. For this, we apply a concept from matrix perturbation theory. Since the previous errors were expressed as the difference in the Frobenius norm, which is the sum of the squared difference of each entry, we see the matrix as having noise introduced. The following lemma provides us with the maximal deviation of each eigenvector in terms of the error in the matrix and the corresponding eigenvalue. The lemma is stated as follow.

\begin{lemma}[Variant of Davis-Kahan sin$\theta$ Theorem \cite{yu2015useful}]\label{SinTheta}
Let $A$ and $\hat A$ be symmetric matrices with eigenvalue $\lambda, \hat \lambda$ and eigenvector $v, \hat v$. Then, 
\begin{align}
    \|\hat v_i - v_i\| \leq \frac{\|A - \hat A\|}{\min (| \hat\lambda_{i-1} - \lambda_i |, |\hat\lambda_{i+1} - \lambda_i|)}
\end{align}
\end{lemma}

Notice that ${\dot Z}$ is not necessarily symmetric. To take advantage of the theorem, we look at ${\dot Z}^T{\dot Z}$ instead. Observe that the eigenvector of ${\dot Z}^T{\dot Z}$ is the right singular vector of ${\dot Z}$, and the eigenvalue is the squared singular value. Furthermore, ${\dot Z}^T{\dot Z}$ is symmetric. The application of this theorem is then possible.

Let $\sigma_i$ here denote the $i$-th singular value of $A$, and $\hat \sigma_i$ denote the $i$-th singular value of $\hat A$. We recall Corollary \ref{alg:PCA}. We know that $|\sigma_i^2 - \hat\sigma_i^2 | \leq \frac{\eta\gamma}{10}\|{\dot Z}\|_F$, and $\sigma_i^2 - \sigma_{i-1}^2\leq \eta_5\|{\dot Z}\|_F^2$. Thus, we can derive 
\begin{align}
    |\hat\sigma_{i+1}^2 - \sigma_i^2 | \leq \eta_5\|{\dot Z}\|_F^2 - \frac{\eta_5\gamma}{10}\|{\dot Z}\|_F
\end{align}
 
This takes care of the denominator of the theorem. The numerator can be bounded by with the same inequality given,
\begin{align}
    \|{\dot Z}^T{\dot Z} - \hat {\dot Z}^T \hat {\dot Z} \| &= \|{\dot Z}^T{\dot Z} - {\dot Z}^T\hat {\dot Z} + {\dot Z}^T\hat {\dot Z} - \hat {\dot Z}^T\hat {\dot Z} \|_F \\
    &\leq \|{\dot Z}^T\|\|{\dot Z}-\hat {\dot Z}\|_F + \|\hat {\dot Z}\|\|{\dot Z} - \hat {\dot Z}\|_F \\
    &= O(E_4\|{\dot Z}\|_F)
\end{align} 

Thus, using Lemma \ref{SinTheta} to bound the perturbation of right singular vectors, to get that \begin{align}
    \|W - \hat W\|_F &\leq \sqrt{J}(\frac{E_4}{\eta_5 \|A\|_F - \eta_5\gamma/10} + \epsilon_5 ) \\ 
    &\leq \sqrt{J}(\frac{E_4}{\eta_5 \|\dot X\|\|X\|^{-1} - \eta_5\gamma/10} + \epsilon_5 )
\end{align}

Lastly, we want to find the error of the output sample. We will generalize it into bounding the matrix product with the Frobenius norm. We get that the final matrix product is bounded by
\begin{align}
    \|ZW - \widehat{ZW} \|_F &\leq \|Z(W - \hat W)\|_F + \|(Z - \hat Z)\hat W \|_F \\
    &\leq E_5 + E_2(1 + E_5)
\end{align}

This concludes the correctness proof. We see that it is possible to pick the parameters in a way that bounds the error to an arbitrary value with high probability. Furthermore, it is suspected that the upper bound can we lowered significantly \cite{tang2019quantum}. 

\subsection{Parameter Selection}\label{subsec:parameter}

In this section, we will use the result from the previous section to suggest a procedure for choosing the parameters of each step to bound the error to an arbitrary $\epsilon \in (0,1)$.

First, we see that the final error is bounded by the expression
\begin{align}
    E_5 + E_2(1 + E_5) = O(E_2 + E_5)
\end{align}

The error resulting from step 2 ($E_2$) is given in equation $\ref{eq:E2}$. 
\begin{align}
    E_2 &\leq \sqrt{d}\epsilon_1(1 + \eta_1\epsilon_1^2) + \epsilon_2 \\
    &= O(\sqrt{d}e_1 + \epsilon_2) 
\end{align}

The error resulting from step 5 ($E_5$) is recursively defined from the errors of the previous steps. Upon expanding it, we get the following.
\begin{align}
    E_5 &= O(\sqrt{J}(\frac{E_4}{\eta_5 \|\dot X\|\|X\|^{-1} - \eta_5\gamma/10} + \epsilon_5 )) \\
    &= O(\sqrt{J}(\frac{\|\dot X\|E_3 + \epsilon_4}{\eta_5 \|\dot X\|\|X\|^{-1} - \eta_5\gamma/10} + \epsilon_5 )) \\
    &= O(\sqrt{J}(\frac{\|\dot X\|(\sqrt{r}\epsilon_1/\theta + (1 + \sqrt{r}\epsilon_1)(2 + \sqrt{r}\epsilon_1)/\theta + \epsilon_3) + \epsilon_4}{\eta_5 \|\dot X\|\|X\|^{-1} - \eta_5\gamma/10} + \epsilon_5 )) \\
    &= O(\sqrt{J}(\frac{\|\dot X\|(r\epsilon_1/\theta + \epsilon_3) + \epsilon_4}{\eta_5 (\|\dot X\|\|X\|^{-1} - \gamma/10)} + \epsilon_5 ))
\end{align}

Furthermore, there is a special parameter selection scheme from implementing Lemma \ref{alg:PCA}. Condensing the above information, we get the following the table below.

\begin{center}
\begin{tabular}{ c|c }

 Parameter & Value (take the minimum of) \\
  \hline
 $\epsilon_1$ & $\frac{\epsilon_1'\|X\|_F^3}{\sigma^3}, \epsilon_1'^2 \eta, \frac{\sigma}{4\|X\|_F^2}$  \\
 $\eta_1$ & $\epsilon_1$  \\ 
 $\epsilon_2$ & $\epsilon$  \\ 
 $\epsilon_3$ & $\frac{\epsilon}{\|\dot X\|_F \sqrt{J}}$  \\ 
 $\epsilon_4$ & $\frac{\epsilon}{\sqrt{J}}$  \\ 
 $\epsilon_5$ & $\frac{\epsilon_5'\|\dot X\|_F^3}{\theta^3\sigma^3}, \epsilon_1'^2 \eta, \frac{\sigma\theta^2}{4\|\dot X\|_F^2}$  \\ 
 $\eta_5$ & $1/(\|\dot X\|\|X\|^{-1} - \gamma/10)$  \\ 
 
\end{tabular}
\end{center}

where $\epsilon_1' = \min\{\frac{\epsilon}{\sqrt{d}}, \frac{\epsilon\sigma^2\theta}{\|X\|_F^2 \|\dot X\| \sqrt{J}}\}$, and $\epsilon_5' = \frac{\epsilon}{\sqrt{J}}$.

\subsection{Alternative Algorithm for Querying}

Algorithm \ref{alg:MainAlgo} outputs a sample from specified row of the desired matrix. We can modify the algorithm to output a specified entry by slightly altering the last step. The algorithm is as follow.

\begin{algorithm}\label{alg:MainAlgo}
\caption{Quantum-Inspired Slow Feature Analysis for Query Access}
\KwIn{data matrix $X^T$, normalized and polynomially expanded, ${\dot{X}^T} \in \R^{d \times n}$, obtained through differentiating $X$}
\KwOut{$Y(j, \cdot)$, for some specified $j$} 
Gain sample and query access to $\hat U^T, \hat V^T, \hat \Sigma\inv$, the approximate SVD of $X$ \\
Gain sample and query access to $\hat Z^T = \hat V \hat U^T$ \\
Gain sample and query access to $\hat {B^{-1/2}} = \hat V(\Sigma^T \Sigma)^{-1/2} V^T \equiv \hat V \hat \Sigma\inv \hat V^T$ \\
Gain sample and query access to the product $\hat {\dot Z} = \dot X \hat{B^{-1/2}}$  \\
Gain sample and query access to right singular values of $\hat A$ \\
Store the lowest $J$ right singular vectors of $\hat {\dot Z}$ in matrix $\hat W$ \\
Output $Y(i,j) = \langle \hat Z(i, \cdot), \hat W(\cdot, j)\rangle $
\end{algorithm}

The modification changes just the last step to outputing the inner-product, which takes $O(d \max \{q(\hat Z), q(\hat W)\})$ time. In this case, one need not to store the $\hat W$ in the data structure, as described in the run time analysis for step 5. Query access from step 4 is sufficient. The argument for the correctness still applies, but the run time is reduced from a quadratic dependence on the dimensions to linear dependence.

If one chooses to obtain sample access to $\hat W$, there is an alternative way of approximating the $(i,j)$-th entry.

\begin{lemma}\label{alg:innerProduct}
For some $x,y \in \mathbb{R}^n$, given $SQ(x)$, $Q(y)$, we can estimate  $\langle x, y \rangle$ to additive error $\epsilon$ and failure probability $\delta$ in query and time complexity \\
$O(\|x\|^2\|y\|^2 \frac{1}{\epsilon} \log\frac{1}{\delta} (s(x) + q(x) + q(y))+n(x))$ 
\end{lemma}

From paying the computational cost for sample access, the lemma above can offer a method for approximating an entry of $Y$ in run time independent of the dimensions, which can be preferable in some scenarios.

This completes the proof of Theorem \ref{eq:MainTheorem}.

% \subsection{Data Centering}

% As mentioned before, Lemma \ref{SinTheta} allows us justify a variety of procedures involving singular value decomposition. Note that we assumed the input data matrix, $X$, to be normalized. Using Lemma \ref{MatrixMult}, we propose the following method for approximately centering the data in $O(n\log n)$ time, given that the unnormalized when entered into the data structure.

% \begin{lemma}
% Given sample and query access to $X \in \R^{n \times d}$, we can normalize $X$ to an additive error $\epsilon$ in time $O(\frac{\|X\|_F}{}$
% \end{lemma}
% \begin{algorithm}
% \caption{Sublinear-time Data Centering}
% \KwIn{Data matrix $X \in \R^{n\times d}$}
% \KwOut{Centered data matrix $\X$}

% Prepare sample and query access to matrix $\mathcal{N} = I - \frac{1}{n}\one$, where $\one$ is the all-1 matrix \\
% Output sample and query access of $\X$ by performing the matrix multiplication $\mathcal{N}X$
% \end{algorithm}

% Note that we can have $SQ(\mathcal{N})$ to be constant time operations following the instructions:
% \begin{itemize}
%     \item $S(\mathcal{N})$ outputs an integer uniformly in the set $\{1, \dots, n\}$
%     \item $S(\mathcal{N}(i, \cdot))$ that outputs $i$ with probability $(\frac{n-1}{n})^2$, and any other index with probability $1/n^2$
%     \item $Q(\mathcal{N})$ outputs $(n-1)/n$ for $\mathcal{N}(i,i)$, $i \in [n]$, and $-1/n$ otherwise.
%     \item $\nu(\mathcal{N})$ outputs $\sqrt{\frac{2n-1}{2}}$
% \end{itemize} 

\section{Conclusion}

In this article, we have shown that we can perform slow feature analysis in time that scales logarithmically to the number of data points, excluding the time complexity of data preprocessing. This reduction in run time complexity implies that the algorithm can be applied onto problems with exponentially larger data sets. The technique used can potentially be used for solving a variety of other problems that involves intensive matrix multiplication or linear algebraic operations. On a broader level, it would be interesting to see the limit of quantum-inspired methods, and if it can act as a replacement for real quantum algorithms. However, currently there exists possible implementation issues for quantum-inspired methods. Arrazola et. al did an experimental study on the computational issues of Tang's recommendation algorithm. \cite{arrazola2019quantum} They observed that the quality of the output significantly relies on condition that the input matrix is low rank with low conditional numbers (the ratio between the largest and smallest singular values). Thus, more research regarding the theoretical properties ought to be conducted prior to implementing the sublinear procedures onto real life data sets.

\section{Acknowledgement}

We thank Professor Ning Xie from Florida International University for helpful comments and discussions. 

\printbibliography

\end{document}